\documentclass[aps,pra,reprint,showpacs,showkeys,superscriptaddress]{revtex4-1}
\usepackage{lipsum}
\usepackage{graphicx}
\usepackage{units}
\usepackage{siunitx}
\usepackage{hyperref}
\usepackage{verbatim,color}

\usepackage{amssymb}
\usepackage{amsmath}
\usepackage{bbm}

\newcommand {\Fig}[1] {Fig.~\ref{#1}}
\newcommand {\Eqn}[1] {Eq.~\ref{#1}}

\newcommand{\CiteVideo}[0] {\cite{[{See Supplemental Material at [URL] for video evidence of how region boundaries were mapped}]empty2}}

\begin{document}

% Use the \preprint command to place your local institutional report
% number in the upper righthand corner of the title page in preprint mode.
% Multiple \preprint commands are allowed.
% Use the 'preprintnumbers' class option to override journal defaults
% to display numbers if necessary
%\preprint{}

%Title of paper
\title{Phase and micromotion of Bose-Einstein condensates in a time-averaged ring trap}

% repeat the \author .. \affiliation  etc. as needed
% \email, \thanks, \homepage, \altaffiliation all apply to the current
% author. Explanatory text should go in the []'s, actual e-mail
% address or url should go in the {}'s for \email and \homepage.
% Please use the appropriate macro foreach each type of information

% \affiliation command applies to all authors since the last
% \affiliation command. The \affiliation command should follow the
% other information
% \affiliation can be followed by \email, \homepage, \thanks as well.
\author{Thomas A. Bell}
\email[]{t.bell4@uq.edu.au}
\affiliation{ARC Centre of Excellence for Engineered Quantum Systems (EQuS), School of Mathematics and Physics, University of Queensland, St Lucia, QLD 4072, Australia}
\author{Guillaume Gauthier}
\affiliation{ARC Centre of Excellence for Engineered Quantum Systems (EQuS), School of Mathematics and Physics, University of Queensland, St Lucia, QLD 4072, Australia}
\author{Tyler W. Neely}
\affiliation{ARC Centre of Excellence for Engineered Quantum Systems (EQuS), School of Mathematics and Physics, University of Queensland, St Lucia, QLD 4072, Australia}
\author{\\Halina Rubinsztein-Dunlop}
\affiliation{ARC Centre of Excellence for Engineered Quantum Systems (EQuS), School of Mathematics and Physics, University of Queensland, St Lucia, QLD 4072, Australia}\
\author{Matthew J. Davis}
\affiliation{ARC Centre of Excellence in Future Low-Energy Electronics Technologies, School of Mathematics and Physics, The University of Queensland, Brisbane QLD 4072, Australia}
\author{Mark A. Baker}
\affiliation{ARC Centre of Excellence for Engineered Quantum Systems (EQuS), School of Mathematics and Physics, University of Queensland, St Lucia, QLD 4072, Australia}

\date{\today}

%%%%%%%%%%%%%%%%%%%%%%%%%%%%%%%%%%%%%%%%%%%%%%%%%%%
%%%%%%%%%%%%%%%%%%%%%%%%%%%%%%%%%%%%%%%%%% Abstract %%%%
%%%%%%%%%%%%%%%%%%%%%%%%%%%%%%%%%%%%%%%%%%%%%%%%%%%
\begin{abstract}
Rapidly scanning magnetic and optical dipole traps have been widely utilised to form time-averaged potentials for ultracold quantum gas experiments.  Here we theoretically and experimentally characterise the dynamic properties of Bose-Einstein condensates in ring-shaped potentials that are formed by scanning an optical dipole beam in a circular trajectory.  We find that unidirectional scanning leads to a non-trivial phase profile of the condensate that can be approximated analytically using the concept of phase imprinting.  While the phase profile is not accessible through in-trap imaging, time-of-flight expansion manifests clear density signatures of an in-trap phase step in the condensate, coincident with the instantaneous position of the scanning beam.  The phase step remains significant even when scanning the beam at frequencies two orders of magnitude larger than the characteristic frequency of the trap.  We map out the phase and density properties of the condensate in the scanning trap, both experimentally and using numerical simulations, and find excellent agreement.  Furthermore, we demonstrate that bidirectional scanning eliminated the phase gradient, rendering the system more suitable for coherent matter wave interferometry.
\end{abstract}

% Insert suggested PACS numbers
% Insert suggested keywords
% 67.85.Hj	Bose-Einstein condensates in optical potentials
% 37.10.Gh	Atom traps & guides
% 67.85.-d	Dynamic properties of condensates; excitations & superfluid flow
\maketitle

%%%%%%%%%%%%%%%%%%%%%%%%%%%%%%%%%%%%%%%%%%%%%%%%%%%
%%%%%%%%%%%%%%%%%%%%%%%%%%%%%%%%%%%%%%%% Introduction %%%%
%%%%%%%%%%%%%%%%%%%%%%%%%%%%%%%%%%%%%%%%%%%%%%%%%%%
\section{Introduction}

The ability to engineer trapping potentials for ultracold quantum gases has enabled their use to  study a wide range of macroscopic quantum phenomena.  In the late 1990s the most commonly used potentials were relatively simple magnetic traps generated by current carrying coils and wires \cite{PhysRevA.58.R2664,PhysRevLett.81.5310}, and optical dipole traps formed at the focus of single laser beams \cite{PhysRevLett.57.314}.  Today increasingly sophisticated traps for ultracold gases employ a range of techniques, including optical lattices \cite{Bloch2005}, combinations of independent lasers \cite{PhysRevLett.74.3577},  rf-dressing of magnetic potentials \cite{Zoway01,Colombe04,Schumm2005}, and pattern projection using spatial light modulators \cite{PhysRevA.73.031402,doi:10.1063/1.5009584} or digital micro mirror devices \cite{1367-2630-14-7-073051,Gauthier:16}.
\vspace{2ex}

An important step toward the first experimental demonstration of Bose-Einstein condensates (BECs) was the development of the Time-Orbiting Potential, or TOP trap~\cite{PhysRevLett.74.3352}.  It had previously been observed that spin polarised atoms confined in a quadrupole magnetic trap could undergo Majorana transitions near the zero of the magnetic field. As the cloud evaporatively cooled, atoms spent more time in the vicinity of the trap minimum, increasing the rate of atom loss \cite{Berg89}. However, it was realised that adding a rapidly rotating bias field could solve this problem. If the field rotation frequency was fast enough that atoms did not move far during a period, but slow enough that the atomic spin could adiabatically follow the local magnetic field direction, then to a good approximation the system experienced a pseudo-static potential, equivalent to the instantaneous potential time-averaged over one field rotation period. This is conceptually similar to rotating electric fields used to confine ions in Paul traps \cite{Berk08}.

TOP traps were a popular choice for many of the early experiments on ultracold gases, however today relatively few groups make use of these potentials.  One of the drawbacks of TOP traps is that the potentials are not actually static; beyond the zeroth order approximation, there exists a micromotion of the trapped atoms.  This was examined in detail experimentally by the Arimondo group in Refs.~\cite{PhysRevLett.85.4454,0953-4075-33-19-320}. Challis \emph {et al.} also presented a detailed theoretical study of micromotion in a TOP trap \cite{Challis04}.

The same principle of time-averaging can be applied to optical dipole traps. If the laser is scanned sufficiently quickly, atoms will experience a Time-Averaged Optical Potential (TAOP) \cite{Schnelle:08,Raman2001,PhysRevA.61.031403,Rudy:01}.  A variety of trapping geometries for degenerate quantum gases have been formed using this method, including line~\cite{Schnelle:08}, ring~\cite{1367-2630-11-4-043030} and lattice~\cite{1367-2630-13-4-043007} potentials. The prospect of atomtronic applications, where analogues of electronic circuits for atoms are engineered, has motivated studies into persistent currents \cite{1367-2630-16-1-013046}, integrated matterwave circuits \cite{1367-2630-17-9-092002}, and Josephson junctions \cite{PhysRevLett.111.205301}. Parallel matterwave splitting \cite{Roberts:142} and efficient runaway optical evaporation \cite{PhysRevA.90.051401} have also been demonstrated.
\pagebreak

%%%%%%%%%%%%%%%%%%%%%%%%%%%%%%%%%%%%%%%%%%%%%%%%%%%
%%%%%%%%%%%%%%%%%%%%%%%%%%%%%%%%%%%%%%%% Figure 1 %%%%
%%%%%%%%%%%%%%%%%%%%%%%%%%%%%%%%%%%%%%%%%%%%%%%%%%%
\begin{figure}[t]
\includegraphics[width=\linewidth, keepaspectratio]{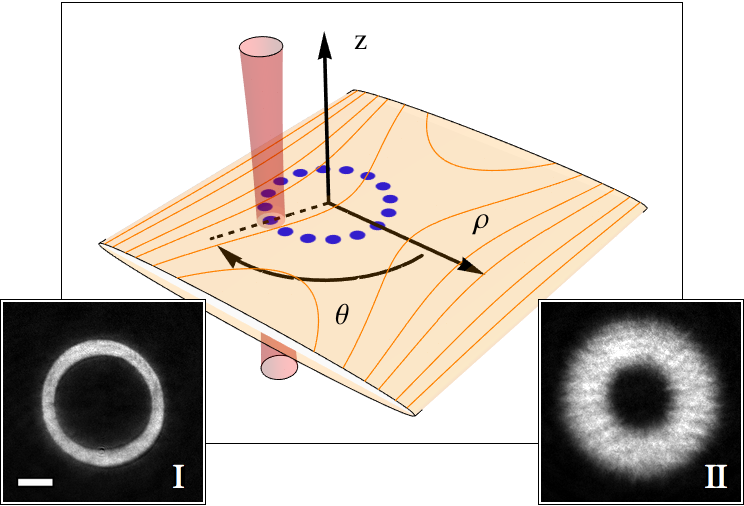}
\caption{\label{fig1}
(Color online).  Schematic of the experimental setup.  The ring potential is formed by combining a red-detuned laser sheet that confines atoms in the $z=0$ plane with a vertically propagating red-detuned beam that is rapidly scanned in the $xy$ plane. (Insets) 
Experimental absorption images of the ring condensate density, following feedforward correction. The images are taken after \SI{1}{\milli\second}~(I) and \SI{20}{\milli\second}~(II) TOF expansion, and re-scaled to the peak density. The ring has radius $R=\SI{82}{\micro\meter}$, and the scale bar in (I) has \SI{50}{\micro\meter} length.
}\end{figure}

Our group has recently studied BECs in time-averaged traps generated by scanning an optical dipole beam using a two-dimensional acousto-optical deflector (2D-AOD), with a particular emphasis on ring-shaped traps~\cite{LargeRing2016}. We have developed a control algorithm that measures the atom distribution in the ring through absorption imaging, then applies an intensity correction to the scanning beam. This feedforward technique reduces the trap depth fluctuations of the ring to less than 10\% RMS of the chemical potential.  While images of the atoms taken in-trap show a smooth atomic density profile, we observe a prominent density feature in the ring, for scan frequencies of the dipole trap less than a few kHz.  The coincidence of this feature with the instantaneous beam location suggests that it results from the scanning motion of the dipole beam, and motivates a more careful study of micromotion in these systems.  In particular, such micromotion would likely affect the BEC phase profile, and therefore be detrimental to matter-wave interferometry.
\vspace{1ex}

In this work, we further characterise the properties of a BEC within a scanned optical ring potential, and identify several features arising from the time dependent nature of the potential. In particular, we observe density and phase features of the BEC that follow the scanned trapping beam; the later we show may be understood using the principle of phase imprinting \cite{Denschlag97}.  While the magnitude of these features can be reduced by increasing the scan frequency, they cannot be eliminated entirely.  Finally, we demonstrate an adapted scan strategy that produces a more uniform azimuthal phase.

%%%%%%%%%%%%%%%%%%%%%%%%%%%%%%%%%%%%%%%%%%%%%%%%%%%
%%%%%%%%%%%%%%%%%%%%%%%%%%%%%%%%%%%%% Experiment Design %%%%
%%%%%%%%%%%%%%%%%%%%%%%%%%%%%%%%%%%%%%%%%%%%%%%%%%%
\section{Experimental setup}
\label{sec:ExpSum}
Our experimental apparatus and procedure have been previously described in Ref.~\cite{LargeRing2016}.  Numerous subsequent improvements are described in Appendix~\ref{sec:ImproveBEC}. Briefly, we form $^{87}\text{Rb}$ Bose-Einstein condensates (BECs) in the ${F=1}$, $m_F=-1$ state, with a typical atom number $N_0\approx2\times10^6$, temperature $T \approx \SI{45}{nK}$, and $75\%$ condensate fraction.  The atoms are confined to a toroidally shaped trap formed by the combination of two red-detuned $\lambda=\SI{1064}{\nano\meter}$ Gaussian beams, illustrated in Fig.~\ref{fig1}.   Vertical harmonic confinement is provided by the cylindrical sheet beam, which propagates along the y-axis and has $1/e^2$ waists of $\sigma_x=\SI{1.25}{\milli\meter}$ and $\sigma_z=\SI{27}{\micro\meter}$.  To create a time-averaged ring trap, the scanning beam increments around $p$ discrete points at radius $R$ in the $xy$ plane. The scanning beam has a waist of $\sigma_\rho=\SI{26.5}{\micro\meter}$. With points spaced by $0.65\, \sigma_{\rho}$, the discrete scan produces rings with smoothly controlled trap depth. The resulting time-averaged ring potential provides harmonic radial confinement.
\vspace{0.8ex}

An acousto-optic deflector (AOD) [IntraAction DTD-274HA6] is used to scan the beam with frequency $f_s$. The AOD access time constrains $f_s\leq\SI{6.25}{\kilo\hertz}$ for our standard ring of radius $R=\SI{82}{\micro\meter}$ and $p=32$ points. Our typical vertical and radial trapping frequencies are $f_z=\SI{140}{\hertz}$ and $f_\rho=\SI{50}{\hertz}$ respectively. The discrete scan affords point-wise control over the local beam power. To improve trap uniformity, the atom distribution at \SI{1}{\milli\second} time-of-flight (TOF) is initially observed, and an iterative feedforward correction to the $p$ beam powers around the ring performed [\Fig{fig1}(I)].  Corrections are subsequently refined using the \SI{20}{\milli\second} TOF distribution, producing ring BECs with uniform density [\Fig{fig1}(II)].

%%%%%%%%%%%%%%%%%%%%%%%%%%%%%%%%%%%%%%%%%%%%%%%%%%%
%%%%%%%%%%%%%%%%%%%%%%%%%%%%%%%%%%%%%%%% Analytics %%%%
%%%%%%%%%%%%%%%%%%%%%%%%%%%%%%%%%%%%%%%%%%%%%%%%%%%
\section{Analytical condensate phase ~~\\~~~in time-averaged ring traps}
\label{sec:analytics}
In this section we develop an analytical description for the phase profile of the condensate held in a time-averaged ring trap. The ring potential is generated by scanning a single Gaussian beam around a circular path of radius $R$, with angular frequency $\omega=2\pi f_s$. When the path curvature may be neglected $(\sigma_\rho \ll R)$, the instantaneous potential can be approximated by
\begin{equation}
V(\rho,\theta, t) = \left(\frac{\overline{V}\sqrt{8\pi}R}{\sigma_\rho}\right) e^{-2\left[\left(\frac{\rho-R}{\sigma_\rho}\right)^2+\left(\frac{R (\theta-\omega t)}{\sigma_\rho}\right)^2 \right]} \,.
\label{eqn:potential}
\end{equation}
The time-averaged trap depth is
\begin{equation}
\overline{V}=-m\,\pi^2\sigma^2_\rho\, f^2_\rho \, ,
\label{eqn:vbar}
\end{equation}
where $m$ is the particle mass,  $\sigma_\rho$ the scan beam waist and $f_\rho$ the time-averged radial trap frequency. The condensate wave function $\Psi(\rho,\theta,t)$ evolves according to the Gross-Pitaevskii equation (GPE)
\begin{equation}
i \hbar \, \frac{\partial \Psi}{\partial t} = \left(\frac{-\hbar^2}{2m} \nabla^2 + V  + g \left|\Psi\right|^2 \right) \Psi \, ,
\label{eqn:GPE}
\end{equation}
where $g=4\pi{\hbar}^2 a_s/m$ is the three dimensional coupling constant, and $a_s$ the s-wave scattering length. For sufficiently large scanning frequency $\omega$, the potential term evolves faster than the timescale over which kinetic and interaction terms evolve. We explore this limit in the following derivation, by neglecting these terms in the GPE, and retaining only the potential term.  Expressing the polar wave function in the Madelung form,
\begin{equation}
\Psi(\rho,\theta,t) = \sqrt{n(\rho,\theta,t)} \,e^{-i\, \phi(\rho,\theta,t)} \, .
\label{eqn:mandelung}
\end{equation}
Substituting \Eqn{eqn:mandelung} into \Eqn{eqn:GPE}, then collecting imaginary terms, 
\begin{equation}
\frac{\partial \,n(\rho,\theta,t)}{\partial t} = 0\,.
\end{equation}
The initial density profile thus remains constant in time within this approximation; we therefore adopt a Thomas-Fermi density for the time-averaged potential \cite{PhysRevA.74.023617}. The Thomas-Fermi radius $R_\rho$ and chemical potential $\mu$ are given by
\begin{equation}
R_\rho= \sqrt{\frac{\mu}{2\,m\,\pi^2 f_\rho^2}} {\quad\mbox{,}\quad} \mu = \sqrt{\frac{2Ng \,m f_\rho f_z}{R}} \, .
\label{eqn:fermi}
\end{equation}
Collecting real terms gives the equation of motion for the condensate phase
\begin{equation}
\hbar \, \frac{\partial \phi (\rho,\theta, t)}{\partial t} = V(\rho,\theta, t)\,,
\label{eqn:imprint}
\end{equation}
which shows the trapping beam continuously imprints the condensate phase while scanning \cite{Denschlag97}. Solving \Eqn{eqn:imprint} using \Eqn{eqn:potential}, the condensate phase is
\begin{equation}
\phi=\frac{\delta_\phi}{2} \left[\kappa-\text{erf}\left(\frac{\sqrt{2}\,R (\theta-\omega t)}{\sigma_\rho}\right)\right] e^{-2\left(\frac{\rho-R}{\sigma_\rho}\right)^2},
\label{eqn:phase}
\end{equation}
where we define the phase step
\begin{eqnarray}
\delta_\phi &=&\left(\frac{\overline{V}}{\hbar f_s}\right)\, .
\label{eqn:deltaPhase}
\end{eqnarray}
The integral constant $\kappa(\rho,\theta,t)$ must satisfy the periodic phase and velocity boundary conditions
\begin{align}
\label{eqn:pBound}
\phi(\rho,\theta+2\pi,t) &= \phi(\rho,\theta,t) + 2\,\pi \,q  \, ,\\
\label{eqn:vBound}
\vec{\nu}(\rho,\theta+2\pi,t) &= \vec{\nu}(\rho,\theta,t) \, .
\end{align}
The condensate velocity is defined by
\begin{eqnarray}
\vec{\nu}\,(\rho,\theta,t) &=&\left(\frac{-\hbar}{m}\right)\, \nabla \phi(\rho,\theta,t) \, ,
\label{eqn:vForm}
\end{eqnarray}
and integer winding number $q$ accounts for a persistent flow. Toward solving these boundary conditions, we use \Eqn{eqn:phase}~\&~\ref{eqn:vForm} to compute the polar velocity field
\begin{align}
\label{eqn:vComp}
\!\! \!\vec{\nu}\,(\rho,&\theta,t) = \left(\frac{\hbar\,\delta_\phi}{2 m}\right)\exp\left(\frac{-2(\rho-R)^2}{\sigma_\rho^2}\right) \times\\
&\quad \left\{ \frac{4\left(\rho-R\right)}{\sigma_\rho^2}\left[ \kappa - \text{erf}\left(\frac{\sqrt{2} R (\theta-\omega t)}{\sigma_\rho}\right) \right] \hat{\rho} \right.\nonumber\\
&\quad \left. + \frac{1}{\rho}\left[ \sqrt{\frac{8}{\pi}}\frac{R}{\sigma_\rho} \exp\left(\frac{-2 R^2 (\theta-\omega t)^2}{\sigma_\rho^2}\right) - \frac{\partial\kappa}{\partial\theta} \right] \hat{\theta} \right\}. \nonumber
\end{align}
Using an integral constant of the form
\begin{equation}
\kappa(\rho,\theta,t) = \left[\frac{1}{\pi}+\left(\frac{2\,q}{\delta_\phi}\right) e^{2\left(\frac{\rho-R}{\sigma_\rho}\right)^2}\right] (\theta-\omega t) + \kappa_0
\label{eqn:kappa}
\end{equation}
satisfies \Eqn{eqn:pBound}~\&~\ref{eqn:vBound} for any constant $\kappa_0$.  While non-linear solutions in $\theta$ are possible, they do not minimise the system energy. We hereon let $q=0$ for simplicity. For the BEC to have zero time-averaged velocity, $\kappa_0=0$. For $\kappa_0$ greater (less) than zero the mean radial velocity monotonically decreases (increases) with displacement $|\rho-R|$.
\vspace{-1ex}

\begin{figure*}[t]
\includegraphics[width=\linewidth, keepaspectratio]{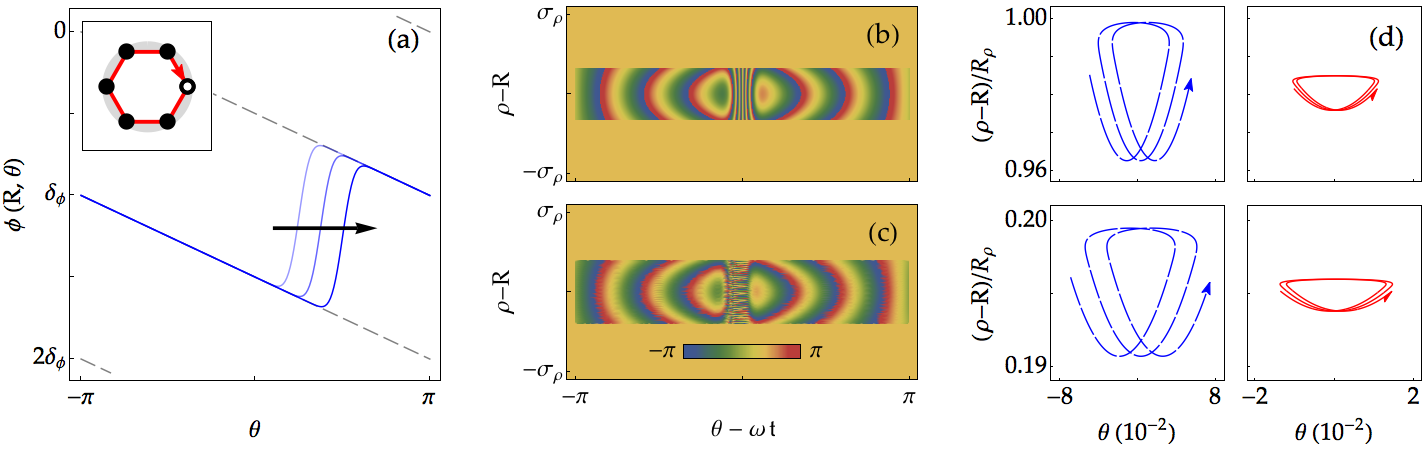}
\caption{\label{fig2}
Analytical solutions for the unidirectionally scanned ring condensate phase. 
(a) Snapshots at three equally spaced times of the azimuthal phase profile of the scanned ring trap, at $\rho=R$. The phase step $\delta_\phi$ advances with the scanning beam location. The black arrow indicates the direction of beam motion. (Inset) Schematic of the unidirectional scan ordering. Black dots represent the discrete scan locations around the ring, commencing at $\theta=0$ (open circle). The red arrow shows the scan order.
(b)~Density plot of the analytical phase solution [\Eqn{eqn:phase}] for the scanning beam at $\theta={\omega} t$, and scan frequency $f_s=\SI{0.5}{\kilo\hertz}$. The phase is shown for $|{\rho-R}|<{R_\rho}$. 
(c) Density plot of the numerical phase solution, for the same parameters as (b) and as discussed in Sec.~\ref{sec:numeric}.
(d) Particle trajectories which illustrate the amplitude of micromotion within the time-averaged trap, for scan frequencies $f_s=\SI{0.5}{\kilo\hertz}$ (blue dashed) and $f_s=\SI{6.25}{\kilo\hertz}$ (red solid). The resulting micromotion is plotted over three scan cycles for initial radial positions $R_\rho$ (top) and $0.2R_\rho$ (bottom). The amplitude of the azimuthal and radial motion reduces at higher scan frequency
}\end{figure*}

The imprinted BEC phase solution described by \Eqn{eqn:phase} is visualised in \Fig{fig2}(a) for $\kappa_0=0$. The profile $\phi(R,\theta)$, shown at three subsequent times, demonstrates the phase step $\delta_\phi$ advances with the trapping beam. The 2D phase profile is shown in \Fig{fig2}(b). Beyond the Thomas-Fermi approximation, the condensate will respond to the associated velocity field, and confined particles will undergo micromotion. Using Euler integration, we solve \Eqn{eqn:vComp} to trace four single particle paths in \Fig{fig2}(d). These open orbits induce net movement around the ring, biasing one propagation direction for interferometric applications \cite{LargeRing2016}. This may serendipitously provide a scan frequency controlled calibration mechanism; the scan direction may be reversed or the frequency changed.

%%%%%%%%%%%%%%%%%%%%%%%%%%%%%%%%%%%%%%%%%%%%%%%%%%%
%%%%%%%%%%%%%%%%%%%%%%%%%%%%%%%%%%%%%%%% Figure 2 %%%%
%%%%%%%%%%%%%%%%%%%%%%%%%%%%%%%%%%%%%%%%%%%%%%%%%%%

%%%%%%%%%%%%%%%%%%%%%%%%%%%%%%%%%%%%%%%%%%%%%%%%%%%
%%%%%%%%%%%%%%%%%%%%%%%%%%%%%%%%%%%%%%%% Numerics %%%%
%%%%%%%%%%%%%%%%%%%%%%%%%%%%%%%%%%%%%%%%%%%%%%%%%%%
\section{Numerical condensate solution \\~in time-averaged ring traps}
\label{sec:numeric}
In this section we use numerical simulations of the GPE to characterise the dynamics of condensates in time-averaged ring traps \cite{Dennis2013}. These results confirm the analytical predictions from Sec.~\ref{sec:analytics}, and provide additional physical insight about the region of stable confinement.
\vspace{1ex} 

The three-dimensional simulation of the Gross-Pitaevskii equation is a numerically demanding task. While it is feasible to find the ground state for a static 3D potential, and even to simulate dynamics of the system following a disturbance, it is extremely challenging to fully simulate a 3D BEC in a scanning trap where there is a significant separation of time scales between the scanning frequency and the typical trap frequencies, or in time-of-flight expansion where a significant spatial domain is required. We have therefore developed an approximate two-dimensional reduction of the GPE to find both the state of the system in the scanning trap, and to simulate the expansion dynamics in time-of-flight. Full details of this methodology will be discussed elsewhere.
\vspace{1ex} 

\begin{figure*}[t]
\includegraphics[width=\linewidth, keepaspectratio]{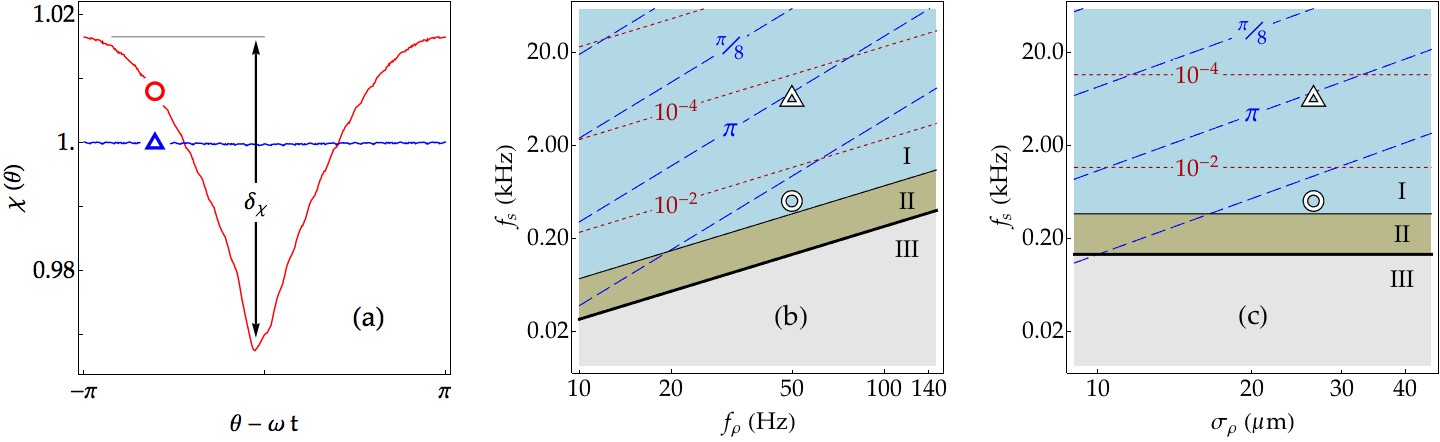}
\caption{\label{fig3} 
Numerical solutions for the unidirectionally scanned ring condensate.
(a) Normalised angular density profiles $\chi(\theta)$ for scan frequencies $f_s=\SI{6.25}{\kilo\hertz}$ (blue triangle) and $f_s=\SI{0.5}{\kilo\hertz}$ (red circle) before time-of-flight expansion. The definition of the density step $\delta_\chi$ is indicated.
(b-c) Characterisation of three distinct regions of BEC behaviour following adiabatic loading into the scanned potential; a function of scan frequency $f_s$, radial trap frequency $f_\rho$, and scan beam waist $\sigma_\rho$. Experimental values quoted in text are fixed for all other parameters. The regions are
(I) Experimentally accessible \emph{confined} region; 
(II) \emph{Transition} region where the time-averaged condition begins to fail;
(III) \emph{Untrapped} region where atoms are numerically unconfined.
Region boundaries are defined where trapped density profiles demonstrate irregularity (I-II), and phase profiles randomise (II-III); see supplementary information \CiteVideo. The contours represent the phase step $\delta_\phi$ (dashed) and density step $\delta_\chi$ (dotted). The white circle and triangle indicate the parameters for the data presented in (a).
}\end{figure*}

The numerical simulations result in condensate phase profiles with identical shape and step $\delta_\phi$ to the analytical result [\Fig{fig2}(b-c)]. However, as the kinetic energy term is now included, the numerical density is non-uniform. We find that the normalised azimuthal density profile
\begin{eqnarray}
\label{eqn:dense}
\chi(\theta,t) &=& \frac{2\pi\, n(R,\theta,t)}{\int^{2\pi}_0 n(R,\theta,t) \,d\theta}
\end{eqnarray}
rotates with the beam and phase profile [\Fig{fig3}(a)]; the minimal density point coincides with the instantaneous beam location. Through numerical data regression, we find the normalised density step is
\begin{equation}
\label{eqn:deltaDense}
\delta_\chi = \gamma \left(\frac{f_\rho}{f_s}\right)^2\,,
\end{equation}
where parameter $\gamma = 5.5\pm 0.6$. The size of the density step intuitively decreases with increasing scan frequency. The phase and density dynamics in the scanned ring are summarised in \Fig{fig3}(b,c). Phase $\delta_\phi$ and density $\delta_\chi$ step contours are shown for variable scan frequency $f_s$, trap frequency $f_{\rho}$ and beam waist $\sigma_{\rho}$. 
\vspace{1ex}

%%%%%%%%%%%%%%%%%%%%%%%%%%%%%%%%%%%%%%%%%%%%%%%%%%%
%%%%%%%%%%%%%%%%%%%%%%%%%%%%%%%%%%%%%%%% Figure 3 %%%%
%%%%%%%%%%%%%%%%%%%%%%%%%%%%%%%%%%%%%%%%%%%%%%%%%%%

From the simulations we have identified three distinct regions of behaviour, as per  \Fig{fig3}(b,c). In Region I, the \emph{confined} region, the scan frequency $f_s$ is sufficiently high that the condensate remains trapped following transfer to the scanning potential, and is well described by the time-averaged potential. In Region II, the \emph{transition} region, as the scan frequency $f_s$ is reduced, the time-averaged condition begins to break down. The numerical simulations show atoms being lost from the scanning potential during the transfer; see supplementary information \CiteVideo. We do not experimentally observe trapping in this region. Finally, as the scanning frequency is further decreased, the system crosses into Region III, the \emph{untrapped} region, where all atoms are lost from the BEC during transfer into the time-averaged ring.  The simulations show the phase step $\delta_{\phi}$, density step $\delta_{\chi}$ and the region boundaries are all independent of $f_z$,  $R$, $g$ and $N_0$.
\vspace{1ex}

Outside region (I), the kinetic energy contribution to the GPE becomes more significant. Given the region boundaries are independent of $g$ and $N_0$, they must be determined by single particle physics.   We therefore define the kinetic energy per particle
\begin{equation}
\label{eqn:KenSingle}
K(\rho,\theta,t) =  \left(\frac{-\hbar^2}{m}\right)\frac{\Psi^{*} \nabla^2 \Psi}{\Psi^{*} \Psi} \, ,
\end{equation}
by numerically dividing the kinetic energy density by the number density. The maximal kinetic energy $K_0$ occurs at the centre of the scanning beam. The (II-III) region boundaries in \Fig{fig3}(b-c) coincide with where $K_0$ equals the time-averaged depth $\overline{V}$ [\Eqn{eqn:vbar}], defining the maximal density step $\delta_{\chi0}\approx1.5$. For $\delta_{\chi}=1.5$, the minimal density in \Fig{fig3}(a) drops to zero. Since the density must remain positive, the limit $\delta_\chi \ll \delta_{\chi0}$ is physically well justified. From \Eqn{eqn:deltaDense}, the minimal scan frequency
\begin{equation}
f_s \gg \sqrt{\frac{\gamma}{\delta_{\chi0}}} f_\rho \approx  2 f_\rho  \,. 
\end{equation}
Radial parametric driving hence induces the loss within region II. Although we have restricted our analysis to a ring geometry, the kinetic energy arguments may provide guidance for determining the scan requirements in other time-averaged configurations, including anharmonic and uniform potentials~\cite{Gauthier:16}. The effect of the scan driven dynamics within TAOP will furthermore have implications for phase sensitive applications such as matter wave interferometry. One approach to mitigate imprinted phase structure is demonstrated in Sec.~VI.
%\vspace{-5ex}

%%%%%%%%%%%%%%%%%%%%%%%%%%%%%%%%%%%%%%%%%%%%%%%%%%%
%%%%%%%%%%%%%%%%%%%%%%%%%%%%%%%%%%%%%%%% Figure 4 %%%%
%%%%%%%%%%%%%%%%%%%%%%%%%%%%%%%%%%%%%%%%%%%%%%%%%%%
\begin{figure*}[t]
\includegraphics[width=\linewidth, keepaspectratio]{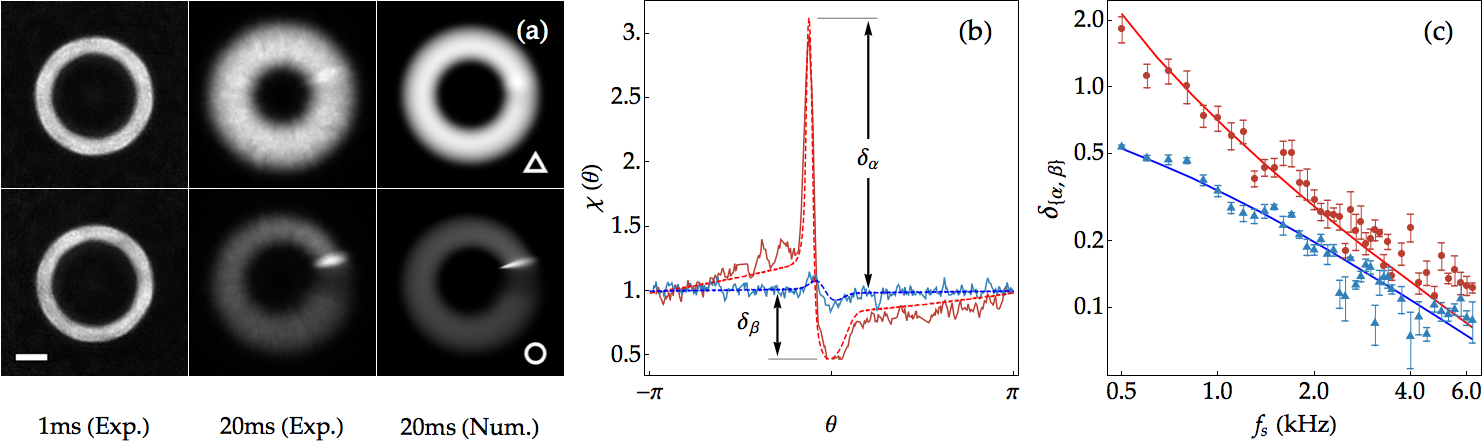}
\caption{\label{fig4} 
Consequences of the condensate phase profile within time-averaged optical ring potentials.
(a) Mean experimental absorption images for scan frequencies $f_s=\SI{6.25}{\kilo\hertz}$ (triangle) and $f_s=\SI{0.5}{\kilo\hertz}$ (circle), taken after $\SI{1}{\milli\second}$ and $\SI{20}{\milli\second}$ time-of-flight expansion.  Imaging noise at $\SI{1}{\milli\second}$ expansion obscures the numerically predicted density step, and the images are experimentally indistinguishable. Following $\SI{20}{\milli\second}$ expansion the change in the condensate for different scan rates becomes apparent, and agrees with the numerical simulation results. The scale bar has $\SI{50}{\micro\meter}$ length.
(b)~Normalised angular density profiles $\chi(\theta)$ for the $\SI{20}{\milli\second}$ TOF images in (a). The profile asymmetry motivates the definition of peak $\delta_\alpha$ and trough $\delta_\beta$ amplitudes. Experimental (solid) and numerical (dashed) profiles are overlaid.
(c) A comparison of experimental density peaks (red circles) and troughs (blue triangles) with simulations (solid lines). The AOD access time limits scan frequency $f_s\le\SI{6.25}{\kilo\hertz}$ for our ring geometry, while beyond $f_s\ge\SI{0.5}{\kilo\hertz}$ the atoms are not confined.
}\end{figure*}

%%%%%%%%%%%%%%%%%%%%%%%%%%%%%%%%%%%%%%%%%%%%%%%%%%%
%%%%%%%%%%%%%%%%%%%%%%%%%%%%%%%%%%%%%% Experiment Data %%%%
%%%%%%%%%%%%%%%%%%%%%%%%%%%%%%%%%%%%%%%%%%%%%%%%%%%
\section{Comparison of numerical \& experimental results}
\label{sec:exp} 
In this section we compare the results of numerical GPE simulations with our experimental observations of condensates in time-averaged traps. Through time-of-flight expansion, condensate phase features transform into density features \cite{PhysRevA.74.061601}, which can be observed using absorption imaging. In our experiment, we use the atom density profile in time-of-flight to correct azimuthal trap depth non-uniformities. This unfortunately complicates the interpretation of single absorption images, as any residual corrugations in the trapping potential cannot be readily distinguished from the density features arising from the phase profile. Our solution is to average a sequence of absorption images taken with an incremented final scanning beam location.  The average of these images accumulates the phase induced feature that coincides with the known final position of the scanning beam. Stationary features are conversely removed through averaging; see Appendix~\ref{sec:ramdi} for complete details.

In \Fig{fig4}(a) we compare experimental images of BECs in time-averaged ring traps, scanned at the minimal and maximal experimental scan frequencies, $f_s=\SI{0.5}{\kilo\hertz}$ and $\SI{6.25}{\kilo\hertz}$ respectively. For these measurements we used $N_0=2\times 10^6$, radius $R=\SI{82}{\micro\meter}$, waist $\sigma_\rho=\SI{26.5}{\micro\meter}$, and radial trap frequency $f_\rho=\SI{50}{\hertz}$. The numerical phase and density steps are $\{\delta_\phi, \delta_\chi\}=\{15.1\pi,\, 5.4\times10^{-2}\}$ and $\{1.2\pi, \,3.5\times10^{-4}\}$ for the scan frequencies respectively. We note the phase step ($\delta_\phi>\pi$) is significant even for scan frequencies two orders of magnitude higher than the trap frequency.

For short time-of-flight (\SI{1}{\milli\second}), the phase structure of the condensate has not yet affected the density, and the images are close to the in-trap density. The density profiles $\chi(\theta)$ for both scan frequencies are uniform, and indistinguishable, as the image noise exceeds the anticipated density step $\delta_\chi$. However, after \SI{20}{\milli\second} time-of-flight, the initial phase profile leads to an appreciable peak in the density coincident with the scanning beam location. While most pronounced for $f_s=\SI{0.5}{\kilo\hertz}$, the feature remains observable at \SI{6.25}{\kilo\hertz}, where $f_s=125 f_\rho$. Our simulations support these observations.

Figure \ref{fig4}(b) compares the normalised numerical and experimental density profiles, for $f_s=\SI{0.5}{\kilo\hertz}$ (red) and $f_s=\SI{6.25}{\kilo\hertz}$ (blue), showing excellent agreement. The density profiles can be characterised using the peak $\delta_\alpha$ and trough $\delta_\beta$ amplitude, as defined in \Fig{fig4}(b). At slow scan frequencies, there is an asymmetry in the time-of-flight density profile ($\delta_\alpha >\delta_\beta$), which reduces at higher scan frequencies. The experimental measurements of these quantities over a broad range of scan frequencies are compared with simulations in \Fig{fig4}(c), and again shows excellent agreement.

%%%%%%%%%%%%%%%%%%%%%%%%%%%%%%%%%%%%%%%%%%%%%%%%%%%
%%%%%%%%%%%%%%%%%%%%%%%%%%%%%%%%%%%%%%%% Raster Scan%%%%
%%%%%%%%%%%%%%%%%%%%%%%%%%%%%%%%%%%%%%%%%%%%%%%%%%%
\section{Bidirectional scanning}
In the above analysis, we have shown that the time-averaged ring trap formed from unidirectional scanning results in a non-uniform phase profile that moves with the scanning beam. While the time-averaged imprinted velocity is vanishing, the scanning beam results in BEC micromotion for all scan frequencies. Furthermore, the phase profile becomes visible in the density through time-of-flight expansion. For most applications a more uniform phase profile would be desirable, and so we modify the scanning protocol to achieve this. 

%%%%%%%%%%%%%%%%%%%%%%%%%%%%%%%%%%%%%%%%%%%%%%%%%%%
%%%%%%%%%%%%%%%%%%%%%%%%%%%%%%%%%%%%%%%% Figure 5 %%%%
%%%%%%%%%%%%%%%%%%%%%%%%%%%%%%%%%%%%%%%%%%%%%%%%%%%
\begin{figure*}[t]
\includegraphics[width=\linewidth, keepaspectratio]{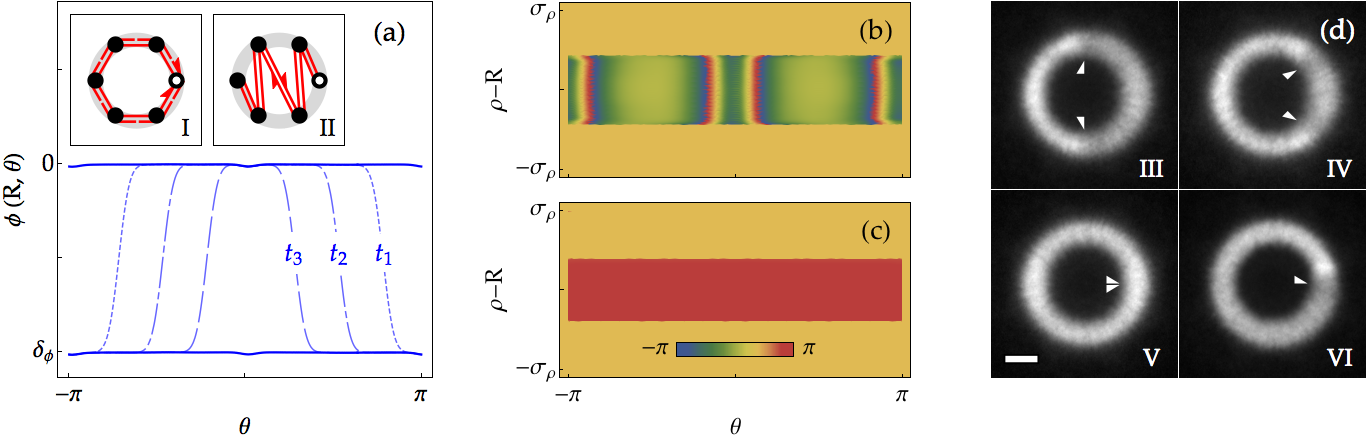}
\caption{\label{fig5} 
Numerical and experimental data for the raster scanned ring condensate.
(a) Snapshots of the central phase profile, for the raster type scan, with $f_s=\SI{6.25}{\kilo\hertz}$. The phase profiles at $\{t_1, t_2, t_3\}$ illustrate times $\{0.25, 0.5, 0.75\}$ through the scan period, showing the azimuthal growth of a flat phase region. The phase profile is uniform at the end of each full scan period. (Inset) (I) Schematic of the idealised \emph{bidirectional} time-averaged scan, with two beams counter-rotating. (II) The \emph{raster} scan ordering approximates the bidirectional scan. Black dots represent the discrete points around the ring, and open circles the starting locations. The red arrows show the scan orders.
(b)~Density plot for the $f_s=\SI{0.7}{\kilo\hertz}$ phase profile after a complete period. The solution is radially truncated where the density becomes vanishing. Residual phase defects, with amplitude $\delta_0$, and located at $\theta=\{0,\pi\}$, result from the temporal offset between the effective scan periods.
(c)~Density plot as in (b) for $f_s=\SI{6.25}{\kilo\hertz}$. The residual phase defects are negligible for our fastest experimental scan speeds.
(d) Experimental density images for $f_s=\SI{0.7}{\kilo\hertz}$ and $\SI{20}{\milli\second}$ TOF. (III-V) Images taken at $\{0.5,0.75,1\}$ through the raster scan period. White arrows indicate the instantaneous beam locations. After one complete period the phase induced density features are absent. (VI)~Comparative density using unidirectional scan ordering. The scale bar in (V) has $\SI{50}{\micro\meter}$ length.
}\end{figure*}

\vspace{1ex}
Rather than unidirectionally scanning a single beam around the ring [\Fig{fig2}(a)], we consider a scheme using two counter-rotating beams. This \emph{bidirectional} scanning is schematically shown in \Fig{fig5}(I). Each scanning beam will individually imprint a phase profile, as described by \Eqn{eqn:phase}, but with opposite gradient. The sum of these profiles results in uniform phase plateaus, aside from at the beam locations. The azimuthal phase is entirely uniform when the beams coincide after each complete scan period.
\vspace{-1ex}

We approximate this bidirectional scheme in our system, using a single beam, by \emph{raster} ordering the scan points, as shown in \Fig{fig5}(II). This ordering alternates between points across the ring, before reversing direction and returning to the initial position. The minimal scan frequency required for the raster ordering must increase since the beam return period is doubled. For our trapping conditions, the raster scan frequency $f_s\ge\SI{0.7}{\kilo\hertz}$. Using $f_s=\SI{6.25}{\kilo\hertz}$, the phase profile in \Fig{fig5}(a) is approximately uniform after each complete period.
\pagebreak

Since the raster ordering shares one beam between two locations, there is a time delay between the counter-rotating periods, equal to the scan point time. This produces two residual phase defects, with amplitude $\delta_0$, and located where the beams pass at $\theta=\{0,\pi\}$. For the scan frequency $f_s=\SI{0.7}{\kilo\hertz}$ these defects in \Fig{fig5}(b) are significant, but may be neglected for $f_s=\SI{6.25}{\kilo\hertz}$ in \Fig{fig5}(c). For both cases the phase step $\delta_\phi$ exceeds the defect $\delta_0$; the relative amplitude $\delta_0/\delta_\phi=0.253$ and $0.013$ respectively.

In \Fig{fig5}(d) we show images of the BEC in the raster scanned trap following \SI{20}{\milli\second} TOF expansion, at three times through the scan period. We find that two density features are visible at the beam locations [\Fig{fig5}(III-IV)]. After a complete raster scan period these features are absent, producing approximately uniform phase and density [\Fig{fig5}(V)]. The density using unidirectional scan ordering shows one rotating phase step [\Fig{fig5}(VI)]. For the raster scan, another saddle-like density artefact manifests through the image processing method developed to extract phase information; see Appendix~\ref{sec:ramdi}. This stationary artefact results purely from the finite response time of the AOD, combined with the raster scan order, and is unrelated to the rotating phase features. We numerically model the BEC expansion from a static potential including this artefact, and find excellent agreement; see Appendix~\ref{sec:saddle}.

We have shown the bidirectional raster scan ordering results in significantly more uniform phase profiles over unidirectional ordering. Combined with our feedforward density correction protocol, the raster ordering produces rings with more uniform density and phase.

%%%%%%%%%%%%%%%%%%%%%%%%%%%%%%%%%%%%%%%%%%%%%%%%%%%
%%%%%%%%%%%%%%%%%%%%%%%%%%%%%%%%%%%%%%%% Conclusion %%%%
%%%%%%%%%%%%%%%%%%%%%%%%%%%%%%%%%%%%%%%%%%%%%%%%%%%
\section{Conclusions}
\vspace{-2ex}
In this paper we have characterised the phase properties and micromotion of Bose-Einstein condensates within ring potentials formed by rapidly scanning an optical dipole potential.  In the limit of rapid scanning, we have shown that condensate dynamics result from local imprinting by the scanned beam; and developed an analytical approximation for the resulting phase profile that is in good agreement with GPE simulations.  We have also derived the scan requirements for confinement within time-averaged potentials. Our kinetic energy arguments result in the often quoted condition for time-averaged traps, that scan frequencies must exceed time-averaged trap frequencies, and inform future investigations into anharmonic time-averaged trapping geometries.
\vspace{2ex}

The phase profile for unidirectional scan ordering results in a clear density feature in time-of-flight that is observable even for scan rates that are much larger than any trapping frequencies.  These phase and density steps are coincident with the instantaneous trapping beam position, and will have consequences for using such potentials as waveguides for coherent matterwaves.  We have therefore subsequently developed a bidirectional raster scan ordering that periodically gives a condensate phase profile that is essentially uniform, making our system more suitable for applications such as atom interferometry.

%%%%%%%%%%%%%%%%%%%%%%%%%%%%%%%%%%%%%%%%%%%%%%%%%%%
%%%%%%%%%%%%%%%%%%%%%%%%%%%%%%%%%%%%% Acknowledgments %%%%
%%%%%%%%%%%%%%%%%%%%%%%%%%%%%%%%%%%%%%%%%%%%%%%%%%%
\begin{acknowledgments}
\vspace{-1ex}
The authors thank Alexander J. Home for technical assistance, and would like to thank Malcolm Boshier for helpful discussions. This research was supported by an Australian Research Council Discovery Grant (DP160102085), and partially supported by the Australian Research Council Centre for Engineered Quantum Systems (project number CE110001013), and the Australian Research Council Centre of Excellence in Future Low-Energy Electronics Technologies (project number CE170100039); and funded by the Australian Government. TAB \& GG acknowledge the support of an Australian Government Research and Training Program Scholarship.
\end{acknowledgments}

%%%%%%%%%%%%%%%%%%%%%%%%%%%%%%%%%%%%%%%%%%%%%%%%%%%
%%%%%%%%%%%%%%%%%%%%%%%%%%%%%%%%%%%%%%%% Appendix 1 %%%%
%%%%%%%%%%%%%%%%%%%%%%%%%%%%%%%%%%%%%%%%%%%%%%%%%%%
\vspace{-2ex}
\appendix
\section{Condensate production}
\label{sec:ImproveBEC}
In our previous work, we observed two deleterious density structures around the ring BEC after time-of-flight (TOF); shown in Fig.~7(b-c) of Ref.~\cite{LargeRing2016}. Density corrugations were invariably observed, accompanied by one larger fringe at the scan beam location. We here refine our experimental procedure to remove the corrugations, thereby enabling our current investigation into the scan dynamics. Our improved experimental method produces greater atom number, and induces fewer excitations during the ring loading and BEC phase transition; producing \Fig{fig1}(I-II) and enabling the research in Sec.~\ref{sec:exp}.

\subparagraph*{\hspace{-2.2ex}Ring Loading:}
Initially we prepare a magneto-optical trap of $^{87}\text{Rb}$ of $N=2\times10^9$ atoms. We then transfer approximately $60\%$ of these to a quadrupole magnetic trap in the hyperfine $F=1$, $m_f=-1$ state, using a field gradient $dB/dz=\SI{1.6}{\tesla /\meter}$. We use RF evaporation to cool the cloud to the temperature $T\approx\SI{5}{\micro\kelvin}$, before lowering the magnetic field gradient to $\SI{0.29}{\tesla /\meter}$ and loading the atoms into a single beam red-detuned dipole trap ($\lambda=\SI{1064}{\nano\meter}$, $\sigma=\SI{65}{\micro\meter}$) \cite{Lin2009}. In our previous work, we evaporated further in this hybrid optical-magnetic trap, and then transferred a BEC into the time-averaged ring potential \cite{LargeRing2016}. In our improved scheme, we halt the evaporation after transfer to the hybrid trap. We simultaneously ramp up the intensity in the optical sheet potential and scanning ring potential, while reducing the single dipole beam power and magnetic field gradient. The edge of the ring overlaps with the position of the cold thermal cloud, and the BEC forms during loading into the ring. We evaporate further by reducing the sheet beam power, to increase the condensate fraction, resulting in BEC of $N_0\approx2\times 10^6$ atoms at a temperature of $T\approx\SI{45}{\nano\kelvin}$, and with $75\%$ condensate fraction. By loading the ring from the thermal cloud, we increase the atom number in the ring considerably, while minimising excitations.
\vspace{2ex}

\subparagraph*{\hspace{-2.2ex}Trap Corrugations:}
We have reduced the spacing of the scan beam points around the ring from $0.7\,\sigma_\rho$ to values below $0.65\,\sigma_\rho$, where $\sigma_\rho$ is the $1/e^2$ waist of the scanning beam. This reduces the depth of corrugations around the ring potential. We note this is less conservative than the figure of $0.527\,\sigma_\rho$ needed for irresolvable points according to the Sparrow criterion \cite{Trypogeorgos:13}. However, any residual corrugations in the condensate density are not visible even after TOF expansion.
\vspace{2ex}

\subparagraph*{\hspace{-2.2ex}Coherence of the BEC:}
The atom number of the ring BECs are sufficiently large that they are not in the phase-fluctuating regime according to the criterion 
\begin{equation}
\label{eqn:coherent}
N_0 > \frac{m \, k_B\, T\,\pi^2 R^2}{\hbar^2} \, ,
\end{equation}
which we derive from Eq.~48 of Ref.~\cite{PhysRevA.82.033607}. Satisfying \Eqn{eqn:coherent} ensures the  coherence length exceeds the farthest separation between two points within the ring condensate, which is half the circumference. For temperature $T = \SI{45}{\nano\kelvin}$, and radius $R=\SI{82}{\micro\meter}$, the ring BEC is fully phase coherent for $N_0>6\times10^5$. 
\vspace{2ex}

\subparagraph*{\hspace{-2.2ex}Improved Feedforward:} As previously reported \cite{LargeRing2016}, we correct imperfections in the ring potential by measuring the atom density distribution from a series of absorption images, and apply iterative corrections to the scanning beam power at each point on the ring.  We previously imaged the ring condensate for $\SI{1}{\milli\second}$ time-of-flight expansion and applied corrections to the beam power inferred from the atom distribution. We have found substantial improvement is possible by using a longer $\SI{20}{\milli\second}$ TOF expansion. The longer expansion time of the ring reduces the absolute optical density, broadens the available image area, and makes residual density corrugations in the trap more apparent, which can then be more accurately corrected. Absorption images are formed on a CCD camera (ProSilica EC1380) with magnification $M=6.38$, and $\SI{1.7}{\micro\meter}$ resolution.

%%%%%%%%%%%%%%%%%%%%%%%%%%%%%%%%%%%%%%%%%%%%%%%%%%%
%%%%%%%%%%%%%%%%%%%%%%%%%%%%%%%%%%%%%%%% Appendix 3 %%%%
%%%%%%%%%%%%%%%%%%%%%%%%%%%%%%%%%%%%%%%%%%%%%%%%%%%

\section{Rotationally Accumulated \\Mean Density Images (RAMDI)}
\label{sec:ramdi}
\begin{figure}[!h]
\includegraphics[width=0.95\linewidth, keepaspectratio]{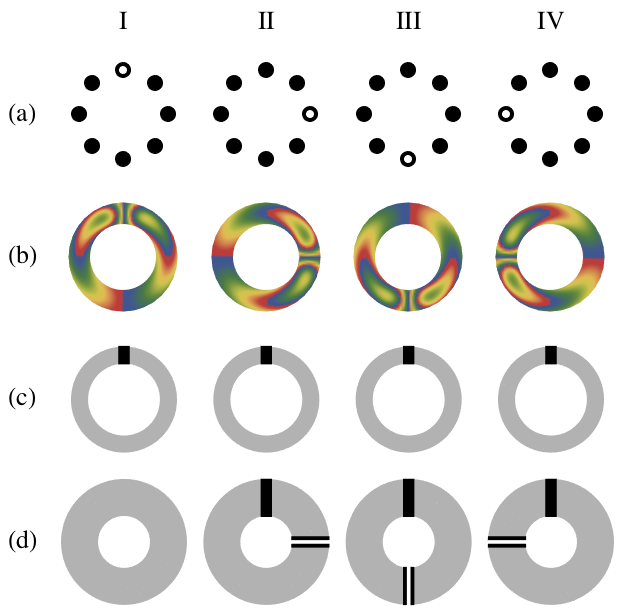}
\caption{\label{fig7}
Schematic of the unidirectional scanning features which motivate the Rotationally Accumulated Mean Density Image (RAMDI) approach.
(a) Black dots represent the discrete points scanned by the trapping beam. The open circle indicates the beam position prior to TOF imaging. Feedforward is performed after TOF and with beam position~(I). For increasing hold times (II-IV), the scan beam position increments around the ring.
(b) Phase profiles in trap.
(c)~Density profiles in trap. Markers indicate the orientation of density features induced through feedforward. These stationary features are stored in the scan beam point powers.
(d) Density profiles after TOF. Markers orientate features resulting from the in-trap phase (striped) and density (solid). Feedforward with beam position (I) induces the density features (c) to correct phase features (b); they consequently cancel for this beam position only.
\vspace{-4ex}
}\end{figure}

Time-of-flight absorption images of the ring BEC may have non-uniform azimuthal density features contributed from two sources. Confined density features, caused by residual trap depth corrugations, produce stationary TOF density features. The imprinted phase profile produces additional TOF density features which rotate with the scanning beam. Our method for extracting only the rotating features is described here. Performing feedforward after expansion adapts the trap depth to compensate both contributions. Static trap depth features are therefore actively introduced by the feedforward algorithm to compensate the rotating features. These features however only overlap and cancel for specific hold times [\Fig{fig7}]. RAMDI are designed to extract only the rotating features, by averaging out the unwanted static features. After performing feedforward at final beam location (I), a series of TOF images are taken with the location incremented through a subset of scan points around the ring (I-IV). In post processing, each image is counter-rotated so the final beam locations coincide before averaging. Scan induced features therefore now constructively add. Stationary features are instead distributed around the ring, and removed through averaging. Mean images also reduce the noise within extracted density profiles. Given our $\SI{20}{\micro\s}$ experimental timing resolution, and $\SI{5}{\micro\s}$ AOD limited point time, eight image series were adopted for $p=32$ rings. 

%%%%%%%%%%%%%%%%%%%%%%%%%%%%%%%%%%%%%%%%%%%%%%%%%%%
%%%%%%%%%%%%%%%%%%%%%%%%%%%%%%%%%%%%%%%% Figure 7 %%%%
%%%%%%%%%%%%%%%%%%%%%%%%%%%%%%%%%%%%%%%%%%%%%%%%%%%

%%%%%%%%%%%%%%%%%%%%%%%%%%%%%%%%%%%%%%%%%%%%%%%%%%%
%%%%%%%%%%%%%%%%%%%%%%%%%%%%%%%%%%%%%%%% Appendix 4 %%%%
%%%%%%%%%%%%%%%%%%%%%%%%%%%%%%%%%%%%%%%%%%%%%%%%%%%
\section{Raster Scan RAMDI Saddle Artefact}
\label{sec:saddle}
Using the RAMDI technique described in Appendix~\ref{sec:ramdi}, we isolate the phase induced TOF density features for time-averaged traps formed with unidirectional ordering.  Unfortunately, for the raster scan, the atomic density based feedforward technique is orientation dependent; the feedforward corrections change for different initial starting locations. When the RAMDI technique is applied to rings with raster scan ordering, there is an additional saddle-like density artefact which cannot readily be isolated from the phase contribution. This arises from the finite response time of the AOD, in conjunction with the raster ordering, and is explained in detail here.
\pagebreak

%%%%%%%%%%%%%%%%%%%%%%%%%%%%%%%%%%%%%%%%%%%%%%%%%%%
%%%%%%%%%%%%%%%%%%%%%%%%%%%%%%%%%%%%%%%% Figure 8 %%%%
%%%%%%%%%%%%%%%%%%%%%%%%%%%%%%%%%%%%%%%%%%%%%%%%%%%
\begin{figure}[t]
\includegraphics[width=0.95\linewidth, keepaspectratio]{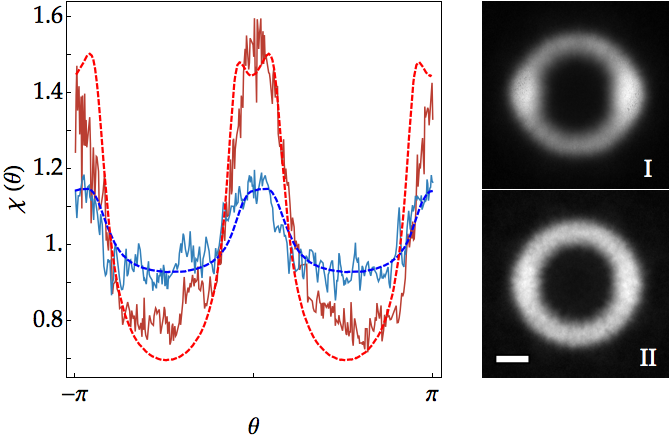}
\caption{\label{fig8}
Diagnostic data for the raster scan ordering RAMDI saddle artefact. Normalised density profiles for experimental RAMDI (solid), taken for \SI{20}{\milli\second} TOF, and scan frequencies $f_s=\SI{1.25}{\kilo\hertz}$ (I) and $f_s=\SI{0.7}{\kilo\hertz}$ (II). The numerical curves (dashed) derived in Appendix \ref{sec:saddle} are overlaid, and assume the phase is uniform. Consistency here indicates that saddling observed experimentally does not result from the residual phase defects $\delta_0$ in \Fig{fig5}(b). The saddle artefact additionally increases with increasing scan frequency, while phase effects should diminish. The scale bar in (II) has \SI{50}{\micro\meter} length.
}\end{figure}

Acousto-optical devices can only transition between discrete sites at a finite rate, constrained by the \emph{access time}; the time for the acoustic wave to travel across the beam waist. The access time is therefore independent of the displacement between sequential points. For a scan consisting $p$ points, $(p-1)$ access times are spent in transit each period. At high scan frequencies $f_s$, where the time per point is comparable to the access time, a non-negligible fraction of the scanned beam power is distributed among the transitions between points. For unidirectional scan ordering, adjacent points are evenly spaced [\Fig{fig2}(a)]. For this geometry the transit power is therefore evenly distributed around the ring, negligibly affecting the azimuthal trap depth.  In the raster scan ordering however, the point-to-point displacement varies through the period [\Fig{fig5}(II)]. The beam transitions therefore non-uniformly contribute to the azimuthal trap depth. This produces a saddle-like trap depth azimuthally around the ring, as effectively more beam power is concentrated at the initial and half period locations, which have the shortest transit displacements.
\pagebreak

To model the effect we assume the beam linearly translates between points during the transition, producing a uniform line potential with Gaussian edges. The fixed AOD access time ensures all connecting potentials have equal integrated power. Our model time-averaged potential then consists $p$ Gaussian potentials and $(p-1)$ overlapping line potentials.  Since the potential is static, the groundstate phase is uniform in trap. The saddle-like TOF density profile, in this case, results from the trapped density fluctuations only [\Fig{fig8}]. Density based feedforward thus does capably remove this modulation. Doing so however makes feedforward scan order dependant, but RAMDI analysis requires that the scan axis incrementally rotate between images. Since performing feedforward for each orientation would defeat the purpose of RAMDI, only the first and middle orientation produce smooth rings. For all other orientations the density is saddled both in trap and through TOF. For high scan frequencies, the time at each discrete point approaches the access time, enhancing the relative contribution of connections to the potential; saddling is more pronounced.
\vspace{29ex}
\pagebreak

%%%%%%%%%%%%%%%%%%%%%%%%%%%%%%%%%%%%%%%%%%%%%%%%%%%
%%%%%%%%%%%%%%%%%%%%%%%%%%%%%%%%%%%%%%% References %%%%
%%%%%%%%%%%%%%%%%%%%%%%%%%%%%%%%%%%%%%%%%%%%%%%%%%%
\bibliography{taop_arXiv.bib}
\vspace{2ex}

\end{document}